\documentclass[twocolumn,superscriptaddress,10pt,showpacs,prl]{revtex4}%

\usepackage{amsfonts}
\usepackage{amsmath}
\usepackage{amssymb}
\usepackage{graphicx}%
\begin{document}

\title{General Description of Mixed State Geometric Phase}
\author{Mingjun Shi}
\email{shmj@ustc.edu.cn}
\affiliation{Department of Modern Physics, University of Science and Technology of China,
Hefei, Anhui, 230026, China}
\author{Jiangfeng Du}
\email{djf@ustc.edu.cn}
\affiliation{Department of Modern Physics,
University of Science and Technology of China, Hefei, Anhui,
230026, China}
\affiliation{Hefei National Laboratory for Physical
Sciences at Microscale, University of Science and Technology of
China, Hefei, Anhui, 230026, China}
\affiliation{Department of
Physics, National University of Singapore, 2 Science Drive 3,
Singapore 117542}

\begin{abstract}
We consider arbitrary mixed state in unitary evolution and provide a
comprehensive description of corresponding geometric phase in which two
different points of view prevailing currently can be unified. Introducing an
ancillary system and considering the purification of given mixed state, we
find that different results of mixed state geometric phase correspond to
different choice of the representation of Hilbert space of the ancilla.
Moreover we demonstrate that in order to obtain Uhlmann's geometric phase it
is not necessary to resort to the unitary evolution of ancilla.

\end{abstract}

\maketitle

The notion of geometric phase was firstly demonstrated in Pancharatnam's study
of the interference of light in different states of
polarization\cite{Pancharatnam}. Afterwards, Berry discovered a non-trivial
phase factor in adiabatic, cyclic and parametric variations of quantal
system\cite{Berry}. This discovery has prompted more attention and more
activities on search for the geometric structure involved in the evolution of
quantum system. Aharonov and Anandan gave the formalism for geometric phase in
the nonadiabtic case\cite{Aharonov and Anandan}. Samuel and Bhandari made the
extension to nonadiabtic and non-cyclic evolution\cite{Samuel and Bhandari}.
Further generalized case of nonadiabtic, non-cyclic and non-unitary evolution
has also been carried out in \cite{Mukunda,Pati}. It should be emphasized that
the above-mentioned considerations refer to the \textit{pure} quantal state.
The definition of geometric phase in mixed state scenario is still an open
question. Uhlmann\cite{Uhlmann} firstly considered this issue within the
mathematical context of purification. In contrast with the abstract formalism
of Uhlmann's definition, Sj\"{o}qvist \textit{et al} presented a more concrete
formalism of geometrical phase for mixed states in the experimental context of
quantum interferometry\cite{Sjoqvist}. Such two definitions------that is,
Uhlmann's and Sj\"{o}qvist's------would result in different geometric phase of
mixed states\cite{Slater}. In other words, they are accordant with each other
only in the case of pure states. It is unsatisfactory in theoretical point of
view to have two different definitions and results for the same thing. On the
other hand, geometric phase has important significance in quantum computation.
Its potential application to fault-tolerant quantum computation has been the
subject of recent investigation\cite{Jones,Zanardi}. There is no conflict in
treating with the pure state geometric phase, but in actual and experimental
situation the issue of mixed states is unpreventable. Unidentical explanation
on the latter would be a trouble for experimentalist.

Recently in \cite{Ericsson}, Ericsson \textit{et al} compared Uhlmann's and
Sj\"{o}qvist's geometric phase for a mixed states undergoing unitary evolution
and concluded that the former depends not only on the geometry of the path of
system alone but also on a constrained bilocal unitary evolution of the
purified entangled state, whereas the latter is essentially the property of
system alone. Nevertheless there is not a consistent definition and
interpretation of mixed state geometric phase so far.

In this paper, we propose a new definition of parallel transport for mixed
state and give a general description of mixed state geometric phase in unitary
evolution, from which Uhlmann's phase and Sj\"{o}qvist's phase can be deduced.
In other words, such two definitions of geometric phase can be embodied in our
statement. In our approach we deal with the mixed state not in view of
spectrum decomposition (i.e., orthogonal decomposition), but by means of
non-orthogonal decomposition. Of course for given mixed state there are
infinite forms of non-orthogonal decomposition. We will seek the specific one
among them and let each pure component undergo parallel transport. The
geometric phase associated with each pure component is simply the total phase
difference during an interval of unitary evolution. Then we think that the
whole ensemble also undergoes parallel transport and the corresponding
geometric phase factor is the sum of weighted geometric phase factor of pure
component with related visibility. To this end, we adopt the usual method to
deal with a mixed state of some quantum system, that is, introducing an
ancilla and considering the purification of the mixed state in a larger
Hilbert space. As a consequence we find that two different choice of the basis
of ancilla Hilbert space will result in Sj\"{o}qvist's and Uhlmann's phase
respectively. Our consideration need not ask for the help of unitary evolution
of ancilla.

Consider a mixed state of some quantum system with $n$\ energy levels. In
$n$-dimensional Hilbert space $\mathcal{H}^{S}$ the\ initial state can be
expressed as%
\begin{equation}
\rho(0)=%
{\displaystyle\sum\nolimits_{j=1}^{n}}
\lambda_{j}\left\vert e_{j}\right\rangle \left\langle e_{j}\right\vert ,
\end{equation}
where $\lambda_{j}$'s\ and $\left\vert e_{j}\right\rangle $'s\ are
respectively the eigenvalues and eigenstates of $\rho(0)$. We choose $\left\{
\left\vert e_{j}\right\rangle \right\}  _{j=1,\cdots,n}$ as the basis of
$\mathcal{H}^{S}$. Some unitary operator $U(t)$\ determines the evolution of
the system, that is%
\begin{equation}
\rho(t)=U(t)\rho(0)U^{\dag}(t).
\end{equation}
$U(t)$ can be represented as $U(t)=\exp\left(  -iHt\right)  $, where $H$ is
the Hamiltonian of the system and assumed to be independent of time for
simplicity (here $\hbar=1$). Introduce an ancillary system. The Hilbert space
describing the ancilla, named as $\mathcal{H}^{A}$, has the same dimension as
$\mathcal{H}^{S}$. The basis of $\mathcal{H}^{A}$\ is labeled by $\left\{
\left\vert f_{j}\right\rangle \right\}  _{j=1,\cdots,n}$. Then the
purification of $\rho(0)$\ is%
\begin{equation}
\left\vert \Psi(0)\right\rangle =%
{\displaystyle\sum\nolimits_{j=1}^{n}}
c_{j}\left\vert e_{j}\right\rangle \otimes\left\vert f_{j}\right\rangle ,
\end{equation}
where $c_{j}$'s can be regarded as real and positive numbers and in fact
$c_{j}=\sqrt{\lambda_{j}}$. Under bilocal unitary transformation $U(t)\otimes
V(t)$, the time evolution of $\left\vert \Psi(0)\right\rangle $\ is%
\begin{equation}
\left\vert \Psi(t)\right\rangle =%
{\displaystyle\sum\nolimits_{j=1}^{n}}
c_{j}U(t)\left\vert e_{j}\right\rangle \otimes V(t)\left\vert f_{j}%
\right\rangle .\label{Psi(t)}%
\end{equation}
Obviously after tracing out ancilla, we always have $\rho(t)=Tr_{A}\left\vert
\Psi(t)\right\rangle \left\langle \Psi(t)\right\vert $ for arbitrary $V(t)$.
Also we express $V(t)$ as $V(t)=\exp\left(  -iKt\right)  $\ where $K$\ is the
time-independent Hamiltonian of ancilla.

Eq.(\ref{Psi(t)}) can be rewritten as%
\begin{equation}
\left\vert \Psi(t)\right\rangle =%
{\displaystyle\sum\nolimits_{j=1}^{n}}
U(t)CV^{T}(t)\left\vert e_{j}\right\rangle \otimes\left\vert f_{j}%
\right\rangle ,\label{Psi(t)-1}%
\end{equation}
where $V^{T}(t)$\ is the transpose of $V(t)$ and $C$ is a diagonal matrix,
i.e., $C=diag[c_{1},c_{2},\cdots,c_{n}]$. Define%
\begin{equation}
|\widetilde{\psi}_{j}(t)\rangle=U(t)CV^{T}(t)\left\vert e_{j}\right\rangle
,\label{psi-tilde(t)}%
\end{equation}
where the symbol "$\sim$" means that $|\widetilde{\psi}_{j}(t)\rangle$'s are
not necessarily orthogonal with one another and are definitely non-normalized.
We have%
\begin{equation}
\left\vert \Psi(t)\right\rangle =%
{\displaystyle\sum\nolimits_{j=1}^{n}}
|\widetilde{\psi}_{j}(t)\rangle\otimes\left\vert f_{j}\right\rangle .
\end{equation}
As for the system, $\rho(t)$\ is expressed as%
\begin{equation}
\rho(t)=%
{\displaystyle\sum\nolimits_{j=0}^{n}}
|\widetilde{\psi}_{j}(t)\rangle\langle\widetilde{\psi}_{j}(t)|.\label{rho(t)}%
\end{equation}
Generally it is a non-orthogonal decomposition. Denote the normalized form of
$|\widetilde{\psi}_{j}(t)\rangle$ as $|\psi_{j}(t)\rangle=\frac{|\widetilde
{\psi}_{j}(t)\rangle}{\langle\widetilde{\psi}_{j}(t)|\widetilde{\psi}%
_{j}(t)\rangle^{1/2}}$. Then $|\psi_{j}(t)\rangle$, as the pure component of
$\rho(t)$,\ appears in the ensemble with the probability $p_{j}(t)=\langle
\widetilde{\psi}_{j}(t)|\widetilde{\psi}_{j}(t)\rangle$. Now we ask such a
question: \textit{Under what condition each }$|\psi_{j}(t)\rangle
$\textit{\ undergoes parallel transport, i.e., }$\langle\psi_{j}(t)|\frac
{d}{dt}|\psi_{j}(t)\rangle=0$\textit{ for }$j=1,2,\cdots,n$\textit{?}

Note that the form of $V(t)$\ is unknown. Usually $p_{j}(t)$\ is not
invariant. So we are above all faced with such a trouble that the pure
components of $\rho(t)$\ come forth with time-dependent weights (or
probabilities). To avoid this trouble we would "see" the system state from the
ancilla in another viewpoint. Formally speaking, we would choose another basis
of $\mathcal{H}^{S}$, say $\left\{  \left\vert g_{j}\right\rangle \right\}  $,
and suppose the relationship between $\left\{  \left\vert f_{j}\right\rangle
\right\}  $\ and $\left\{  \left\vert g_{j}\right\rangle \right\}  $\ is a
time-independent unitary transformation $Z$ which will be determined later,
that is, $\left\vert f_{j}\right\rangle =Z\left\vert g_{j}\right\rangle $.
Rewrite (\ref{Psi(t)-1}).%
\begin{align}
\left\vert \Psi(t)\right\rangle  & =%
{\displaystyle\sum\nolimits_{j=1}^{n}}
U(t)CV^{T}(t)\left\vert e_{j}\right\rangle \otimes Z\left\vert g_{j}%
\right\rangle \nonumber\\
& =%
{\displaystyle\sum\nolimits_{j=1}^{n}}
U(t)CV^{T}(t)Z^{T}\left\vert e_{j}\right\rangle \otimes\left\vert
g_{j}\right\rangle \nonumber\\
& =%
{\displaystyle\sum\nolimits_{j=1}^{n}}
|\widetilde{\varphi}_{j}(t)\rangle\otimes\left\vert g_{j}\right\rangle
,\label{Psi(t)-2}%
\end{align}
where $|\widetilde{\varphi}_{j}(t)\rangle\equiv U(t)CV^{T}(t)Z^{T}\left\vert
e_{j}\right\rangle $ Then $\rho(t)=%
{\displaystyle\sum\nolimits_{j=0}^{n}}
|\widetilde{\varphi}_{j}(t)\rangle\langle\widetilde{\varphi}_{j}(t)|.$ Now we
hope every $\langle\widetilde{\varphi}_{j}(t)|\widetilde{\varphi}%
_{j}(t)\rangle$ is invariant. It is equivalent to say that all diagonal
elements of matrix $Z^{\ast}V^{\ast}C^{2}V^{T}Z^{T}$ are time-independent.
Note that $C$ is time-independent. If $Z^{\ast}$\ can diagonalize $V^{\ast}$,
i.e., $Z^{\ast}V^{\ast}Z^{T}$ is diagonal (or equivalently $ZKZ^{\dag}$\ is
diagonal), then our hope is realized. Now suppose that we have the
time-independent unitary $Z$ such that%
\begin{align*}
\overline{K} &  \equiv Z^{\ast}K^{T}Z^{T}=ZKZ^{\dag}\\
&  =diag[\kappa_{1},\kappa_{1},\cdots,\kappa_{n}],
\end{align*}
where $\kappa_{j}$'s are all real numbers and in fact the eigenvalues of $K$.
Correspondingly $V(t)$ is transformed to%
\begin{align*}
\overline{V}(t)  & =ZV(t)Z^{\dag}\\
& =diag[e^{-i\kappa_{1}},e^{-i\kappa_{2}},\cdots,e^{-i\kappa_{n}}].
\end{align*}
\ Then we can say that given arbitrary non-diagonal $V(t)$ (or Hamiltonian
$K$) on the basis $\left\{  \left\vert f_{j}\right\rangle \right\}  $\ we can
choose as new basis $\left\{  \left\vert g_{j}\right\rangle \right\}  $ of
$\mathcal{H}^{S}$\ the eigenstates of $K$\ so that in the view of $\left\{
\left\vert g_{j}\right\rangle \right\}  $\ the weight of\ each pure component
of the system mixed ensemble is invariant. This invariant weight is presented
as%
\begin{equation}
q_{j}=\langle\widetilde{\varphi}_{j}(t)|\widetilde{\varphi}_{j}(t)\rangle
=\langle e_{j}|Z^{\ast}C^{2}Z^{T}|e_{j}\rangle.\label{qj}%
\end{equation}
Let $|\varphi_{j}(t)\rangle=q_{j}^{-1/2}|\widetilde{\varphi}_{j}(t)\rangle$.
We have $\rho(t)=%
{\displaystyle\sum\nolimits_{j=1}^{n}}
q_{j}^{2}|\varphi_{j}(t)\rangle\langle\varphi_{j}(t)|$. It is still a
non-orthogonal decomposition of $\rho(t)$.

Next, what about the exact form of Hamiltonian $K$ such that guarantees
$\langle\varphi_{j}(t)|\frac{d}{dt}|\varphi_{j}(t)\rangle=\langle
\widetilde{\varphi}_{j}(t)|\frac{d}{dt}|\widetilde{\varphi}_{j}(t)\rangle=0$
for each $j$? That is, we further hope%
\begin{widetext}%
\begin{gather}
\left\langle e_{j}\right\vert [Z^{\ast}V^{\ast}(t)CU^{\dag}(t)][\overset
{\cdot}{U}(t)CV^{T}(t)Z^{T}+U(t)C\overset{\cdot}{V^{T}}(t)Z^{T}]\left\vert
e_{j}\right\rangle =0,\\
j=1,2,\cdots,n,\nonumber
\end{gather}%
\end{widetext}%
or equivalently%
\begin{gather}
\left\langle e_{j}\right\vert Z^{\ast}V^{\ast}(t)[CHC\nonumber\\
+C^{2}K^{T}]V^{T}(t)Z^{T}\left\vert e_{j}\right\rangle
=0,\label{Condition for K}%
\end{gather}
for $j=1,2,\cdots,n$. Remind ourselves that $Z^{\ast}$ is time-independent and
can diagonalize $K^{T}$ and $V^{\ast}$. We suppose $K^{T}$\ is determined by
the following equation.%
\begin{equation}
C^{2}K^{T}+K^{T}C^{2}=-2CHC.\label{Condition for K -1}%
\end{equation}
Then (\ref{Condition for K}) equals%
\begin{align*}
&  \left\langle e_{j}\right\vert Z^{\ast}\left(  CHC+C^{2}K^{T}\right)
Z^{T}\left\vert e_{j}\right\rangle \\
&  =\frac{1}{2}\left\langle e_{j}\right\vert Z^{\ast}[\left(  CHC+C^{2}%
K^{T}\right)  \\
&  +\left(  -CHC-K^{T}C^{2}\right)  ]Z^{T}\left\vert e_{j}\right\rangle \\
&  =\frac{1}{2}\left\langle e_{j}\right\vert Z^{\ast}\left(  C^{2}K^{T}%
-K^{T}C^{2}\right)  Z^{T}\left\vert e_{j}\right\rangle \\
&  =0.
\end{align*}
That implies $\langle\varphi_{j}(t)|\frac{d}{dt}|\varphi_{j}(t)\rangle=0$.
Thus we have answered the above-mentioned question, that is, first the
Hamiltonian of ancilla has to meet condition (\ref{Condition for K -1}), and
then the basis of Hilbert space of ancilla is such that the Hamiltonian has
diagonal form. We regard this description as the parallel transport of mixed
state in unitary evolution.

The geometric phase of $|\varphi_{j}(t)\rangle$\ is%
\begin{align}
\gamma_{j}(t)  & =\arg\langle\varphi_{j}(0)|\varphi_{j}(t)\rangle\nonumber\\
& =\arg\left\langle e_{j}\right\vert Z^{\ast}CU(t)CV^{T}(t)Z^{T}|e_{j}%
\rangle\nonumber\\
& =\arg\left\langle e_{j}\right\vert Z^{\ast}CU(t)CZ^{T}\overline{V}%
^{T}(t)|e_{j}\rangle\nonumber\\
& =\arg\left\langle e_{j}\right\vert Z^{\ast}CU(t)CZ^{T}|e_{j}\rangle
-\kappa_{j}.\label{gamma(t)}%
\end{align}
The visibility is decided by $\nu_{j}=|\langle\varphi_{j}(0)|\varphi
_{j}(t)\rangle|$.

Retrospecting (\ref{Psi(t)-2}), we can say that $\left\vert \Psi
(t)\right\rangle $\ undergoes parallel transport, and hence give the geometric
phase of $\left\vert \Psi(t)\right\rangle $ as follows.%
\begin{align}
\Gamma(t)  & =\arg\langle\Psi(0)\left\vert \Psi(t)\right\rangle \nonumber\\
& =\arg%
{\displaystyle\sum\nolimits_{j=1}^{n}}
\langle\widetilde{\varphi}_{j}(0)|\widetilde{\varphi}_{j}(t)\rangle
\nonumber\\
& =\arg%
{\displaystyle\sum\nolimits_{j=1}^{n}}
q_{j}\nu_{j}e^{i\gamma_{j}}\label{Gamma(t)}%
\end{align}
Rewrite (\ref{Gamma(t)}) as $e^{i\Gamma(t)}=%
{\displaystyle\sum\nolimits_{j=1}^{n}}
q_{j}\nu_{j}e^{i\gamma_{j}}$ and consider the mixed ensemble of
system. (\ref{Gamma(t)}) is in actually the sum of weighted
geometric phase factor of (non-orthogonal) pure component with the
associated visibility. We regard (\ref{Gamma(t)}) as the geometric
phase of mixed state. Obviously it is gauge-invariant. We would
rather review (\ref{Gamma(t)}) from another point of view.

Note that $\rho(0)=CC^{\dag}$. $\left\vert \Psi(t)\right\rangle $ being\ in
bilocal unitary transformation $U(t)\otimes V(t)$,\ we have $C(t)=U(t)CV^{T}%
(t)$. It is evident that $\rho(t)=C(t)C^{\dag}(t)$. Then $C(t)$\ is the
purification of $\rho(t)$\ in Uhlmann's sense. Let $W(t)=C(t)Z^{T}$. $W(t)$ is
another Uhlmann's purification of $\rho(t)$.\ Furthermore\ we have%
\begin{align*}
&  W^{\dag}(t)\overset{\cdot}{W}(t)\\
&  =Z^{\ast}C^{\dag}(t)\overset{\cdot}{C}(t)Z^{T}\\
&  =-iZ^{\ast}V^{\ast}(t)[CHC+C^{2}K^{T}]V^{T}(t)Z^{T}\\
&  =iZ^{\ast}V^{\ast}(t)[CHC+K^{T}C^{2}]V^{T}(t)Z^{T}\\
&  =W(t)\overset{\cdot}{W^{\dag}}(t),
\end{align*}
where we have used (\ref{Condition for K -1}). Thus $W(t)$\ satisfies
Uhlmann's parallel transport condition, and we obtain Uhlmann's geometric
phase of mixed state.%
\begin{align}
\gamma^{Uhl}(t) &  =\arg Tr\left(  W^{\dag}(0)W(t)\right)  \nonumber\\
&  =\arg Tr[Z^{\ast}CC(t)Z^{T}]\nonumber\\
&  =\arg Tr[CC(t)].
\end{align}
It is straightforward to check $\gamma^{Uhl}(t)=\Gamma(t)$. So our statement
in fact results in Uhlmann's geometric phase. In brief, after choosing the
Hamiltonian of ancilla to meet (\ref{Condition for K -1}), and selecting the
specific basis of $\mathcal{H}^{S}$ to diagonalize such Hamiltonian, we indeed
answer the question\ and meanwhile obtain Uhlmann's phase.

Let's consider a special case in which $V(t)$ is already diagonal in the
original basis $\left\{  \left\vert f_{j}\right\rangle \right\}  $.\ Recall
(\ref{psi-tilde(t)}) and (\ref{rho(t)}). It is easy to see that $p_{j}%
(t)=\langle\widetilde{\psi}_{j}(t)|\widetilde{\psi}_{j}(t)\rangle$\ is
invariant and in fact is $\lambda_{j}$. Define the Hamiltonian of ancilla as
$K_{kl}=-\delta_{kl}H_{kl}$, where $K_{kl}$ and$\ H_{kl}$'s are respectively
the matrix elements of $K$ and$\ H$. The evolution of ancilla is given by
$V(t)=diag[e^{iH_{11}t},e^{iH_{22}t},\cdots,e^{iH_{nn}t}]$. We have
$\langle\widetilde{\psi}_{j}(t)|\frac{d}{dt}|\widetilde{\psi}_{j}(t)\rangle
=0$. Similar to (\ref{Gamma(t)}), the geometric phase of $\left\vert
\Psi(t)\right\rangle $\ in this case is%
\begin{align}
\Gamma^{\prime}(t) &  =\arg\langle\Psi(0)|\Psi(t)\rangle\nonumber\\
&  =\arg%
{\displaystyle\sum\nolimits_{j=1}^{n}}
\langle\widetilde{\psi}_{j}(0)|\widetilde{\psi}_{j}(t)\rangle\nonumber\\
&  =\arg%
{\displaystyle\sum\nolimits_{j=1}^{n}}
\lambda_{j}\left\vert \langle e_{j}|U(t)|e_{j}\rangle\right\vert
e^{i\gamma_{j}^{\prime}(t)},
\end{align}
where $\gamma_{j}^{\prime}(t)=\arg\langle\widetilde{\psi}_{j}(0)|\widetilde
{\psi}_{j}(t)\rangle$. This is just the mixed state geometric phase based on
Sj\"{o}qvist's formalism\cite{Sjoqvist}.

Thus for mixed states our proposal includes the origination of Sj\"{o}qvist's
phase and Uhlmann's phase. We almost get a comprehensive interpretation on
mixed state geometric phase. A remained discomfort in the above discussion is
the time-evolution of the ancilla, i.e., $V(t)$. In the following we try to
remove $V(t)$\ and give a more concise expression.

Actually in the above discussion our purpose of applying $V(t)$ to the
ancilla\ is to cancel out the dynamical phase of the system, or more strictly
speaking, to cancel out the dynamical phase of each component of the mixed
ensemble $\rho(t)$. Then we can re-explain our procedure in alternative
viewpoint as follows. We firstly \textit{postulate} that in some
representation of $\mathcal{H}^{A}$, say $\left\{  \left\vert f_{j}%
\right\rangle \right\}  $, the ancilla is subjected to the Hamiltonian $K$
satisfying (\ref{Condition for K -1}). Then choose another representation
$\left\{  \left\vert g_{j}\right\rangle \right\}  $ of $\mathcal{H}^{A}$ in
which $K$ is diagonal. As stated above, unitary $Z$ transforms $\left\{
\left\vert g_{j}\right\rangle \right\}  $ to $\left\{  \left\vert
f_{j}\right\rangle \right\}  $. Afterwards we can \textit{disregard} $K$ and
the corresponding time evolution operator $V(t)=\exp\left(  -iKt\right)  $. So
$\left\vert \Psi(t)\right\rangle $ is decided as the following.%
\[
\left\vert \Psi(t)\right\rangle =%
{\displaystyle\sum\nolimits_{j=1}^{n}}
U(t)CZ^{T}\left\vert e_{j}\right\rangle \otimes\left\vert g_{j}\right\rangle .
\]
Let
$\vert$%
$\widetilde{\chi}_{j}(0)\rangle=CZ^{T}\left\vert e_{j}\right\rangle $ and\
$\vert$%
$\widetilde{\chi}_{j}(t)\rangle=U(t)$%
$\vert$%
$\widetilde{\chi}_{j}(0)\rangle$. Apparently $\langle\widetilde{\chi}%
_{j}(t)|\widetilde{\chi}_{j}(t)\rangle$ is invariant and is just $q_{j}%
$\ (\ref{qj}). Let $|\chi_{j}(t)\rangle=q_{j}^{-1/2}|\widetilde{\chi}%
_{j}(t)\rangle$. We have%
\begin{align}
\rho(t) &  =%
{\displaystyle\sum\nolimits_{j=0}^{n}}
|\widetilde{\chi}_{j}(t)\rangle\langle\widetilde{\chi}_{j}(t)|\nonumber\\
&  =%
{\displaystyle\sum\nolimits_{j=0}^{n}}
q_{j}|\chi_{j}(t)\rangle\langle\chi_{j}(t)|.\label{rho in khi}%
\end{align}
Also $|\chi_{j}(t)\rangle$'s are not orthogonal with one another and do not
endure parallel transport.\ Let's calculate the dynamical phase of
$\vert$%
$\chi_{j}(t)\rangle$.%
\begin{align*}
&  \phi_{j}^{d}(t)\\
&  =\frac{-i}{q_{j}}%
{\displaystyle\int\nolimits_{0}^{t}}
\langle\widetilde{\chi}_{j}(t^{\prime})|\frac{d}{dt^{\prime}}\left\vert
\widetilde{\chi}_{j}(t^{\prime})\right\rangle dt^{\prime}\\
&  =-\frac{1}{q_{j}}\left\langle e_{j}\right\vert Z^{\ast}CHCZ^{T}\left\vert
e_{j}\right\rangle t
\end{align*}
If applying (\ref{Condition for K -1}), we have $\phi_{j}^{d}(t)=\left\langle
e_{j}\right\vert Z^{\ast}K^{T}Z^{T}\left\vert e_{j}\right\rangle t=\kappa
_{j}t$. The total phase is%
\begin{align*}
\phi_{j}^{t}(t) &  =\arg\langle\chi_{j}(0)|\chi_{j}(t)\rangle\\
&  =\arg\langle e_{j}|Z^{\ast}CU(t)CZ^{T}|e_{j}\rangle.
\end{align*}
Then the geometric phase associated with each
$\vert$%
$\chi_{j}(t)\rangle$ is $\phi_{j}^{g}(t)=\phi_{j}^{t}(t)-\phi_{j}^{d}(t)$.
Apparently $\phi_{j}^{g}(t)=\gamma_{j}(t)$ (see (\ref{gamma(t)})). Hence we
have demonstrated that in order to obtain Uhlmann's phase we may not resort to
the evolution of ancillary system, and all we have to do is to find specific
representation of $\mathcal{H}^{A}$ in which the hermitian matrix $K$
satisfying (\ref{Condition for K -1}) is diagonal.

Conclusively, in the formalism proposed in this paper, the mixed state
geometric phase is provided with unified interpretation. The key point is
Eq.(\ref{Condition for K -1}) which implies the representation transformation
of the ancilla. As stated before, the unitary evolution of ancilla is not
necessary, and it is the choice of representation of ancillary Hilbert space
that results in particular geometric phase. For given mixed state and its
purification, different choice of the representation of $\mathcal{H}^{A}$\ can
be considered as different kind of measurement. So the mixed state geometric
phase behaves somewhat like the result of quantum measurement: Having nothing
to do with $\mathcal{H}^{A}$,\ we would get Sj\"{o}qvist's result; shifting to
the other basis of $\mathcal{H}^{A}$, we would get Uhlmann's result. The
hidden significance of such basis transformation is not clear. We hope our
discussion would be helpful to further research of geometric phase.

Financial support from National Natural Science Foundation of
China (Grant No. 10375057 and No. 10425524), the National
Fundamental Research Program (2001CB309300) and the ASTAR (Grant
No. 012-104-305) is gratefully acknowledged.

\end{document}